\documentclass{elsart}
\usepackage{graphicx,amssymb}
\journal{Physics Letters A}
\begin{document}
\begin{frontmatter}
\title{Calculating adiabatic evolution of the perturbed DNLS/MNLS solitons}
\thanks{This work was supported by the National Natural Science Foundation of China
under Grant No. 10375027 and the SRF for ROCS, SEM of China.}
\author{Xiang-Jun Chen\corauthref{xjc}}
\corauth[xjc]{Corresponding author.}
\ead{xiangjun-chen@21cn.com}
\author{Wa Kun Lam}
\ead{wakunlam@21cn.com}
\address{Department of Physics, Jinan University, Guangzhou 510632, P. R. China}
\begin{abstract}
A symbolic computation technique is developed to calculate
adiabatic evolution equations for parameters of the perturbed
DNLS/MNLS solitons obtained by the recently developed direct
perturbation theory [X.-J. Chen and J. Yang, Phys. Rev. E {\bf
65}, 066608(2002)]. Effects of the intrapulse Raman scattering,
third-order group velocity dispersion, and narrow-banded filters
with nonlinear gain on the MNLS solitons are studied as examples.
\end{abstract}
\begin{keyword}DNLS/MNLS solitons \sep perturbation \sep symbolic computation
\PACS 05.45.Yv \sep 52.35.Bj \sep 42.81.Dp
\end{keyword}
\end{frontmatter}
\section{Introduction}
It has been well known that exactly integrable nonlinear differential
equations support soliton solutions which travel stationarily and collide
elastically. It was also known that the physical situations giving rise to exactly
solvable equations are highly idealized. Perturbations violating
their integrabilities actually exist. If these perturbations are small enough, their
influence on solitons can still be known analytically
by perturbation theories
for solitons(see, e.g., \cite{KaupNewell,Karpman,Kaup90}).
In the picture depicted by perturbation theories, the lowest approximation is an adiabatic
solution in which the soliton keep its profile unchanged while its parameters such as
amplitude, velocity, initial center and initial phase may evolve
adiabatically\cite{KaupNewell,Karpman}. Perturbations can also induce non-adiabatic changes
such as radiation emissions. A mathematically complete perturbation theory can obtain not only
evolution equations for parameters of solitons analytically but also a formula for calculating
the perturbation-induced radiation emission.

With vanishing boundary conditions, $u\to 0$ as $|x|\to\infty$,
the derivative nonlinear Schr\"odinger(DNLS) equation,
\begin{equation}\label{DNLSeq}
iu_t+u_{xx}+i(|u|^2u)_x=0,
\end{equation}
is an integrable model describing small amplitude nonlinear
Alfv\'en waves in a low-$\beta$ (the ratio of kinetic to magnetic
pressure) plasma, propagating strictly parallel to the ambient
magnetic field. Here $u$, $x$, and $t$ represent the transverse
complex magnetic field, the normalized longitudinal coordinate,
and the normalized time, respectively and the subscripts denote
partial derivatives\cite{Rogister71,Mjolhus76,Mio76}.

The modified nonlinear Schr\"odinger (MNLS) equation,
\begin{equation}\label{MNLSeq}
iq_z+\frac12 q_{tt}+is(|q|^2q)_t+|q|^2q=0,
\end{equation}
is an integrable model in describing propagation of femtosecond
pulses in single mode fibers
\cite{Tzoar,Anderson83,Ohkuma87,Doktorov}. For picosecond pulses,
it is well known that the nonlinear Schr\"odinger (NLS, the case
when $s=0$) equation is a good model. The MNLS model includes the
nonlinear dispersion term(the third term on the left) which is one
of the several higher order effects becoming more significant in
the femtosecond region\cite{Agrawal01,Hasegawa95}. Here $u$, $z$,
$t$, and $s$ denote the normalized electric field, the normalized
distance, the normalized time measured in the frame of reference
moving with the pulse, and the relative amplitude of the nonlinear
dispersion, respectively. We assume $s\ge0$ in this letter because
the case for $s<0$ can be obtained by a simple transformation
$t\to-t$. The DNLS equation and the MNLS equation are connected by
a gauge-like transformation(see, e.g. \cite{Ichikawa80}). The
soliton solution for the DNLS with vanishing boundary conditions
was found by the inverse scattering transform
technique\cite{Kaup78} and the soliton solution for the MNLS can
be found by the gauge-like transformation.

Perturbation theory for the DNLS/MNLS solitons was first developed by a method based on the
Riemann-Hilbert problem\cite{Valery98}, which had minor errors in evolution equations for the
initial center and phase and gave no correction beyond the adiabatic approximation. Later a
direct perturbation method\cite{Chen02} corrected these errors and obtained a formula for
calculating the perturbation-induced radiation emission. With well established perturbation
theories, influence of higher order effects and methods of controlling soliton shapes and
frequency shift against perturbations can be studied. However, evolution equations for the
perturbed MNLS solitons are much more complicated than those for the perturbed NLS
solitons\cite{KaupNewell,Karpman}. Problems solved by perturbation theories for DNLS/MNLS
solitons\cite{Valery98,Chen02} were limited to some relatively simple cases so
far\cite{Valery98,Chen02,Afanasev98}. Symbolic technique seems helpful for further studies. In
this letter, we develop a symbolic technique, which is effective for usual perturbations, to
calculate adiabatic evolution equations for parameters of perturbed DNLS/MNLS solitons
automatically. In the MNLS model, only one higher order effect beyond the NLS model, the
nonlinear dispersion, is considered. Other higher order effects, such as the intrapulse Raman
scattering and the third-order dispersion, should be treated as perturbations. Evolution
equations for parameters of the MNLS soliton under some control method, such as narrow-banded
filter and nonlinear gain, will be valuable in estimating these methods. In section 2, we
rewrite those evolution equations obtained in Ref.\cite{Chen02} in slightly simpler forms. In
section 3, we describe the symbolic technique. In section 4, we calculate evolution equations
of the MNLS solitons under intrapulse Raman scattering with local approximation, the
third-order group velocity(GVD) dispersion, and narrow band filters with nonlinear gain. In
small $s$ limit, all these results approach those for the NLS solitons in the literature, as
generally shown in Ref.\cite{Chen02}.
\section{Adiabatic evolution of parameters of the perturbed DNLS/MNLS solitons}
\subsection{Perturbed DNLS solitons}
The DNLS one-soliton\cite{Kaup78} can be rewritten as
\begin{equation}
u_{s}(x,t) = -2\Delta\sin(2\beta)\frac{\cosh(\theta-i\beta)}
{\cosh^2(\theta+i\beta)}\exp(-i\varphi),
\end{equation}
where
\begin{equation}
\theta(x,t)=2\eta(x+4\xi t-x_0),
\end{equation}
\begin{equation}
\varphi(x,t)=2\frac\xi\eta\theta-4(\xi^2+\eta^2)t+\varphi_0,
\end{equation}
\begin{equation}
\xi=\Delta^2\cos(2\beta),\quad \eta=\Delta^2\sin(2\beta), \quad 0<\beta<\pi/2.
\end{equation}
Its amplitude is
\begin{equation}
A=4\Delta\sin\beta.
\end{equation}
Simple calculations yield its energy,
\begin{equation}
E=\int_{-\infty}^{+\infty}dx|u_s|^2=8\beta,
\end{equation}
and the FWHM width,
\begin{equation}
w=\frac1{\Delta^2\sin(2\beta)}\ln\left(\cos\beta+\sqrt{1+\cos^2\beta}
\right).
\end{equation}
In presence of perturbations, the zero on the left hand side of Eq.~(\ref{DNLSeq}) should
be replaced with the perturbation function $ir(u)$.
As $\beta$ is proportional to the energy, considering evolution of $\beta$ and $\Delta$,
instead of $\xi$ and $\eta$ in Ref.~\cite{Chen02}, is more convenient. Evolution equations for
parameters of the DNLS solitons in Ref.~(\cite{Chen02}) are reformulated as,
\begin{equation}
\frac{d\beta}{dt}=\frac1{16\eta}\int_{-\infty}^{\infty}d\theta R_+(\theta,t),
\end{equation}
\begin{equation}
\frac{d\Delta}{dt}=i\frac\Delta{16\eta}
\int_{-\infty}^{+\infty}d\theta \tanh(\theta+i\beta)R_+(\theta,t),
\end{equation}
\begin{equation}
\frac{dx_0}{dt}=\frac{\Delta^2}{16\eta^3}\int_{-\infty}^{+\infty}d\theta\theta\frac
{\cosh(\theta+i3\beta)}{\cosh(\theta+i\beta)}R_-(\theta,t)
-\frac{i}{32\eta^2}\int_{-\infty}^{\infty}d\theta R_-(\theta,t),
\end{equation}
\begin{equation}
\frac{d\varphi_0}{dt}=\frac{\Delta^4}{8\eta^3}\int_{-\infty}^{+\infty}d\theta\theta
R_-(\theta,t)+i\frac{\Delta^2}{16\eta^2}\int_{-\infty}^{+\infty}d\theta
\frac{\cosh(\theta-i\beta)}{\cosh(\theta+i\beta)}R_-(\theta,t),
\end{equation}
where
\begin{equation} \nonumber \label{RpmD}
R_{\pm}(\theta,t) =\overline{u_0(\theta)}[
r_0(\theta,t)\pm\overline{r_0(-\theta,t)}],
\end{equation}
\begin{equation}
u_0(\theta) =u_s\exp(i\varphi), \quad
r_0(\theta,t)=r\exp(i\varphi),
\end{equation}
and the bar stands for complex conjugate in this letter.
\subsection{Perturbed MNLS solitons}
The MNLS one-soliton is
\begin{equation}
q_{s}(t,z) = -2\Delta\sin(2\beta)\frac{\cosh(\theta-i\beta)}
{\cosh^2(\theta+i\beta)}\exp(-i\varphi),
\end{equation}
where
\begin{equation}
\theta(t,z)=4s\eta(t-t_0-z/v),
\end{equation}
\begin{equation}
\varphi(t,z)=2\frac{\xi}{\eta}\theta-8s^2(\xi^2+\eta^2)z-\frac{t}{s}+\frac{z}{2s^2}+
\varphi_0,
\end{equation}
\begin{equation}
\xi=\Delta^2\cos(2\beta),\quad \eta=\Delta^2\sin(2\beta), \quad 0<\beta<\pi/2,
\end{equation}
and $v=s/(1-4s^2\xi)$ is its velocity. Similar to the DNLS soliton, we have
amplitude,
\begin{equation}
A=4\Delta\sin\beta,
\end{equation}
energy,
\begin{equation}
E=\int_{-\infty}^{+\infty}dt|q_s|^2=\frac{4\beta}{s},
\end{equation}
and the FWHM width,
\begin{equation}
\tau=\frac1{2s\Delta^2\sin(2\beta)}\ln\left(\cos\beta+\sqrt{1+\cos^2\beta}
\right),
\end{equation}
of the MNLS soliton. Also, in presence of perturbations, the zero
on the right hand side of Eq.~(\ref{MNLSeq}) should be replaced
with the perturbation function $ir(q)$. Evolution equations for
MNLS soliton parameters in Ref.~\cite{Chen02} are reformulated as,
\begin{equation}\label{dgamma}
\frac{d\beta}{dz}=\frac1{16\eta}\int_{-\infty}^{+\infty}d\theta R_+(\theta,z),
\end{equation}
\begin{equation}\label{dDelta}
\frac{d\Delta}{dz}=i\frac\Delta{16\eta}
\int_{-\infty}^{+\infty}d\theta \tanh(\theta+i\beta)R_+(\theta,z),
\end{equation}
\begin{equation}\label{dt0}
\frac{dt_0}{dz}=\frac{\Delta^2}{32s\eta^3}\int_{-\infty}^{+\infty}d\theta\theta
\frac{\cosh(\theta+i3\beta)}{\cosh(\theta+i\beta)} R_-(\theta,z)
-\frac i{64s\eta^2}\int_{-\infty}^{+\infty}d\theta R_-(\theta,z),
\end{equation}
\begin{equation}\label{dphi0}
\frac{d\varphi_0}{dz}=\frac{\Delta^4}{8\eta^3}
\int_{-\infty}^{+\infty}d\theta\theta R_-(\theta,z)
+i\frac{\Delta^2}{16\eta^2}\int_{-\infty}^{+\infty}d\theta\frac{\cosh(\theta-i\beta)}
{\cosh(\theta+i\beta)}R_-(\theta,z),
\end{equation}
where
\begin{equation} \nonumber
R_{\pm}(\theta,z) =\overline{q_0(\theta)}[ r_0(\theta,z)\pm\overline{r_0(-\theta,z)}],
\end{equation}
\begin{equation}
q_0(\theta) =q_s\exp(i\varphi), \quad r_0(\theta,z)=r\exp(i\varphi).
\end{equation}
\section{A symbolic technique to calculate integrals in evolution equations for soliton parameters}
In general, integrals in evolution equations for perturbed DNLS/MNLS solitons are
rather complicated. Even modern symbolic softwares can not always tackle them directly.
However, taking the perturbed MNLS solitons as an example, most perturbation
functions can be expressed as
\begin{equation}\label{pform}
r(q)=\sum_{k=0}^nr_k(|q|^2)\partial_t^kq,
\end{equation}
where $r_k(|q|^2)$ are complex functions\cite{Agrawal01,Hasegawa95}.
In what follows we will show that for this category of perturbation functions
integrals in the evolution equations can be
systematically solved by the technique of residue theorem.

For perturbation functions in category of Eq.~(\ref{pform}), there are only four types of
integrands in the evolution equations
needed to be tackled. Having been continuated to the whole complex plane of $\theta$, they
are
\begin{enumerate}
\item $f(\theta)$ with $f(\theta+i\pi)=-f(\theta)$,
\item $g(\theta)$ with $g(\theta+i\pi)=g(\theta)$,
\item $F(\theta)=\theta f(\theta)$,
\item $G(\theta)=\theta g(\theta)$.
\end{enumerate}
Within the closed path shown in Fig.1, all of them have and only have a
pair of singularities $\theta_{\pm}=i(\pi/2\pm\beta)$.
Integrations of them on the two vertical line
segments are obviously zero. Using the residue
theorem on the closed path shown in Fig. 1, we get
\begin{equation}
\int_{-\infty}^{\infty}f(\theta)d\theta=i\pi\{\mbox{Res}[f(\theta_+)] +\mbox{Res}[f(\theta_-)]\},
\end{equation}
and
\begin{equation}
\int_{-\infty}^{\infty}F(\theta)d\theta=-i\frac\pi2\int_{-\infty}^{\infty}f(\theta)d\theta +
i\pi\{\mbox{Res}[F(\theta_+)] +\mbox{Res}[F(\theta_-)]\}.
\end{equation}
For $g(\theta)$ and $G(\theta)$, by introducing two auxiliary functions,
\begin{equation}
h(\theta,\rho)=\exp(i\rho\theta)g(\theta),
\end{equation}
and
\begin{equation}
H(\theta,\rho)=\exp(i\rho\theta)G(\theta),
\end{equation}
in which $\rho>0$, we also have
\begin{equation}
\int_{-\infty}^{\infty}g(\theta)d\theta=i2\pi\lim_{\rho\to0^+}\frac{\mbox{Res}[h(\theta_+,\rho)] +
\mbox{Res}[h(\theta_-,\rho)]}{ 1-\exp(-\rho\pi)},
\end{equation}
\begin{eqnarray}\nonumber
\int_{-\infty}^{\infty}G(\theta)d\theta &=&\lim_{\rho\to0^+}\left\{-\pi^2\frac{\mbox{Res}
[h(\theta_+,\rho)] +\mbox{Res}[h(\theta_-,\rho)]}{2\sinh^2(\rho\pi/2)}\right. \\
& &+\left.i2\pi\frac {\mbox{Res}[H(\theta_+,\rho)] +\mbox{Res}[H(\theta_-,\rho)]}
{ 1-\exp(-\rho\pi)}\right\}.
\end{eqnarray}
Therefore, despite the fact that these integrals may not be tackled directly,
calculation of them comes down
to calculation of residues and limitations which can always be done symbolically by
modern commercial mathematical softwares.
\section{Examples of perturbed MNLS solitons}
\subsection{Intrapulse Raman scattering}
In local approximation, the intrapulse Raman scattering was
described by a perturbation function\cite{Agrawal01,Hasegawa95},
\begin{equation}
r(q) = -\sigma_R q(|q|^2)_t,
\end{equation}
where $\sigma_R$ is a constant. We get $R_+=2\bar{q}_0r_0$,
$R_-=0$. This problem is so simple that we don't need the symbolic
technique in the preceding section. Direct integrations yields
\begin{equation}
\frac{dt_0}{dz}=0, \quad \frac{d\varphi_0}{dz}=0,
\end{equation}
\begin{equation}
\frac{d\beta}{dz}=0,
\end{equation}
\begin{equation}
\frac{d\Delta}{dz}=-\frac{32}{3}s\sigma_R \Delta^5(2 + \cos^2
2\beta - 6\beta\cot2\beta)
\end{equation}
It is obvious that $d\Delta/dz$ monotonically decreases for all $\beta\in(0, \pi/2)$. The MNLS
soliton perturbed by the intrapulse Raman scattering retain some similarities with the NLS
soliton. They both have no shift in initial position and phase. Their energies are not
perturbed. They all have frequency redshifts(but in different rates). However, the MNLS soliton
has a decrease in amplitude $A$ and an increase in width $\tau$ while keeping its energy
unperturbed.
\subsection{Third-order GVD dispersion}
The third-order GVD dispersion\cite{Agrawal01,Hasegawa95} is
described by
\begin{equation}
r(q)=\alpha\partial_t^3q,
\end{equation}
where $\alpha$ represents its strength. We have $R_+=0$,
$R_-=2\bar{q}_0r_0$,
\begin{equation} \label{paraEE1}
\frac{d\beta}{dz}=0, \quad \frac{d\Delta}{dz}=0,
\end{equation}
and, using the technique developed in the preceding section,
\begin{eqnarray}\nonumber\label{paraEE3}
\frac{dt_0}{dz} &=&\alpha[3s^{-2} -12(\beta\eta + 4\xi -\beta\xi^2\eta^{-1}) \\
& &+ 16(13\eta^2 + \beta\eta\xi - 15\xi^2 +
9\beta\xi^3\eta^{-1})s^2],
\end{eqnarray}
\begin{eqnarray}\label{paraEE4}\nonumber
\frac{d\varphi_0}{dz} &=& \alpha[-2s^{-3} +12(\beta\eta+4\xi-12\beta\xi^2\eta^{-1})s^{-1}\\
\nonumber
& &+32(-11\eta^2+\beta\eta\xi+3\xi^2-3\beta\xi^3\eta^{-1})s\\
& &+64(5\beta\eta^3 -16\eta^2\xi+14\beta\eta\xi^2
-16\xi^3+9\beta\xi^4\eta^{-1})s^3].
\end{eqnarray}
Within adiabatic approximation, influences of the third-order GVD dispersion on the MNLS
soliton is similar to those on the NLS soliton: the amplitude, width, and the main velocity $v$
are unchanged while inducing shifts on $t_0$ and $\varphi_0$. Numerical simulation showed that
the third-order GVD dispersion induces radiation emission from the NLS soliton\cite{PKAWaiOL}
and the problem needs to be solved by a perturbation theory beyond all order\cite{PKAWai}. With
a similar simulation, one can find similar radiation emission from the MNLS soliton. This means
that the problem may also need a perturbation theory beyond all order.
\subsection{Narrow band filters with nonlinear gain}
For NLS solitons, periodic insertion of narrow band filters was shown to be effective in
reducing the frequency shift of solitons. But it was found that filters may induce background
instabilities. M. Matsumoto et. al. suggested to utilize nonlinear gains to suppress such
instabilities\cite{Matsumoto95}. For further studies, we calculate the adiabatic evolution
equations of a MNLS soliton controlled by both of filters and nonlinear gains here. The
corresponding perturbation function is
\begin{equation}
r(q) =\Gamma q + \kappa\frac{\partial^2 q}{\partial t^2} +
\rho_1|q|^2q + \rho_2|q|^4q,
\end{equation}
where $\Gamma$ is an excess gain, $\kappa>0$ represents the strength of the filter, $\rho_1$
and $\rho_2$ are nonlinear gain coefficients. We get $R_+=2\bar{q}_0r_0$, $R_-=0$, and, with
the symbolic technique in the preceding section,
\begin{eqnarray}
\frac{dt_0}{dz} = 0, \hspace{6mm} \frac{d\varphi_0}{dz} = 0,
\end{eqnarray}
\begin{eqnarray} \nonumber
\frac{d\beta}{dz}&=& -2s^{-2}\kappa\beta -32s^2\kappa(5\beta\eta^2
- 4\eta\xi + 9\beta\xi^2)  \\ \nonumber &&+2\left[\beta\Gamma
+12\kappa\eta - 16\kappa\beta\xi + 4\rho_1(\eta-2\beta\xi)\right. \\
&&\left.+16\rho_2(2\beta\eta^2-3\eta\xi + 6\beta\xi^2)\right],
\end{eqnarray}
\begin{eqnarray} \nonumber
\frac{d\Delta}{dz}&=& \frac{\kappa\Delta(\eta-2\beta\xi)}{s^2\eta}
+\frac{16s^2\kappa\Delta(3\eta^3 -14\beta\eta^2\xi + 11\eta\xi^2 - 22\beta\xi^3)}{\eta}\\
\nonumber &&- \frac{\Delta}{3\eta} \left[3\eta(\Gamma +
16\kappa\beta\eta) - 3\xi(2\beta\Gamma + 24\kappa\eta) +
144\kappa\beta\xi^2 \right.\\ \nonumber
&&+12\rho_1(-3\eta\xi +6\beta\xi^2+ 2\beta\eta^2) \\
&& +\left.32\rho_2(4\eta^3 - 18\beta\eta^2\xi + 15\eta\xi^2 -
30\beta\xi^3)\right].
\end{eqnarray}
\section{Summary and discussion}
In this letter, we develop a symbolic technique to calculate the
adiabatic evolution of perturbed MNLS solitons. Evolution
equations under intrapulse Raman scattering, third-order GVD
dispersion and narrow band filters with nonlinear gains are
calculated with the technique. As $s\to 0$, to keep parameters in
the MNLS soliton physically meaningful, $\xi$ and $\eta$ must be
\begin{equation}
\xi \to \frac{\mu}{2s} + \frac1{4s^2}, \quad \eta \to
\frac{\nu}{2s}.
\end{equation}
The MNLS soliton approach the NLS soliton,
\begin{equation}
q_s(t,z) \to -2\nu\,\mbox{sech}\theta\exp(-i\varphi),
\end{equation}
where
\begin{equation}
\theta=2\nu(t-\hat{t}), \quad \hat{t}=-2\mu z + t_0,
\end{equation}
\begin{equation}
\varphi=4\mu(t-\hat{t})+\hat{\varphi}, \quad \hat{\varphi}=
-4(\mu^2+\nu^2)z + \varphi_0.
\end{equation}
In Ref.\cite{Chen02}, it was generally shown that evolution equations for perturbed MNLS
solitons approach those for perturbed NLS solitons in small $s$ limits. Expanding results in
the preceding section near $s=0$, one can find they do approach their corresponding results for
NLS soliton in the literature\cite{Agrawal01,Hasegawa95,Matsumoto95}.

\begin{figure}[h]
       \centering
        \includegraphics[width=0.6\textwidth]{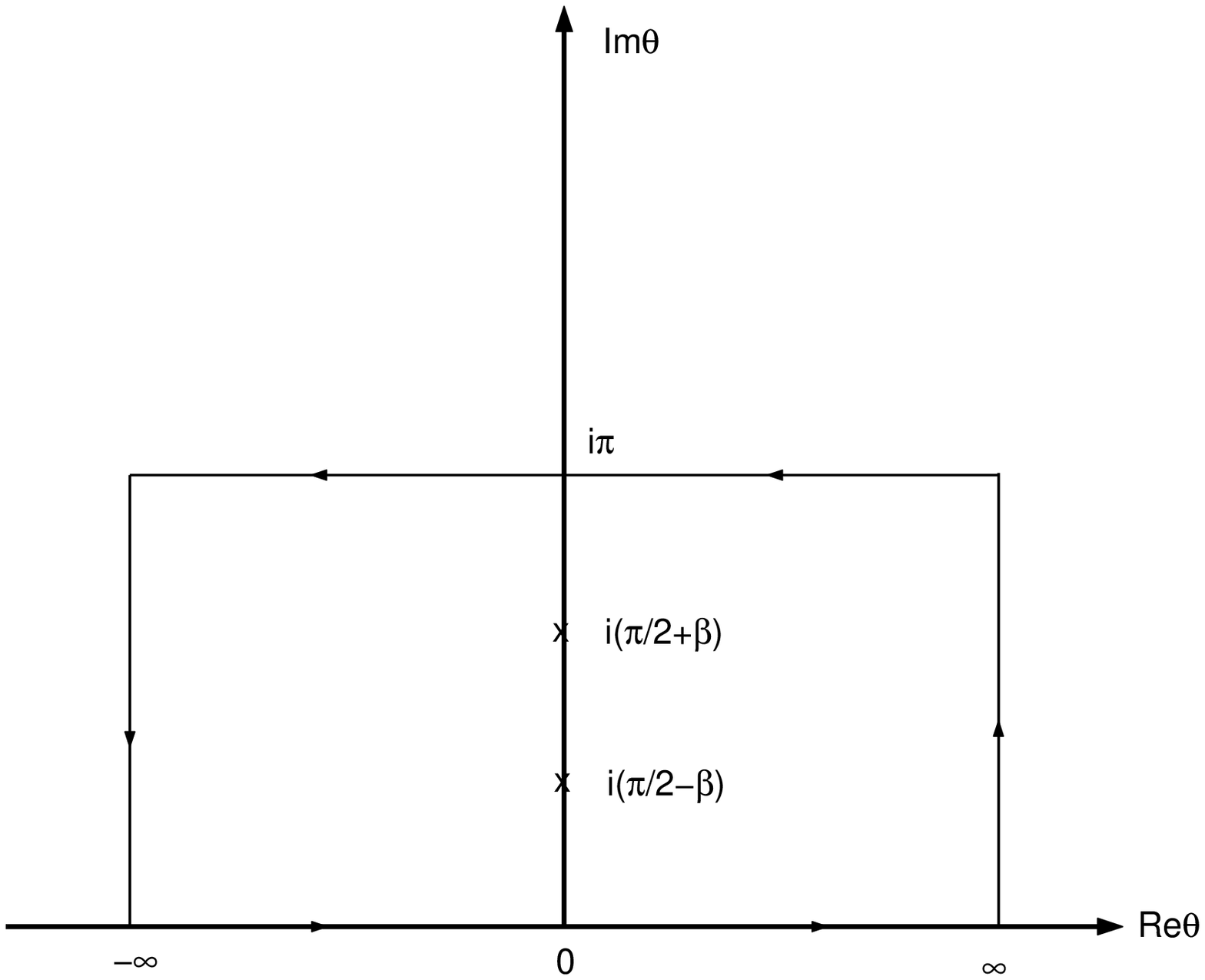}
         \caption{The path for integrals in section 3. The crosses are the singularities.}
         \label{intpath}
  \end{figure}
\end{document}